\documentclass[aps,prl,twocolumn,showpacs,floatfix,nobibnotes,superscriptaddress,amsmath,amssymb,longbibliography]{revtex4-1}

\usepackage{graphicx}
\usepackage{epsfig}
\usepackage{bm}
\usepackage{color}
\usepackage{float}
\usepackage{dcolumn}
\usepackage{multirow} 
\usepackage{hyperref}

\graphicspath{{.}}

\newcommand{\beq}{\begin{equation}}
\newcommand{\eeq}{\end{equation}}
\newcommand{\beqa}{\begin{eqnarray}}
\newcommand{\eeqa}{\end{eqnarray}}

\begin{document}

\title{Structure of the lightest tin isotopes}

\author{T.~D.~Morris}
\affiliation{Department of Physics and Astronomy, University of Tennessee,
Knoxville, TN 37996, USA} 
\affiliation{Physics Division, Oak Ridge National Laboratory,
Oak Ridge, TN 37831, USA} 

\author{J.~Simonis}
\affiliation{Institut f\"ur Kernphysik, TU Darmstadt, Schlossgartenstr. 2, 64289 Darmstadt, Germany} 
\affiliation{ExtreMe Matter Institute EMMI, GSI Helmholtzzentrum f\"ur Schwerionenforschung GmbH, 64291 Darmstadt, Germany}

\author{S.~R.~Stroberg}
\affiliation{TRIUMF 4004 Wesbrook Mall, Vancouver BC V6T 2A3, Canada} 
\affiliation{Physics Department, Reed College, Portland OR, 97202, USA}

\author{C.~Stumpf}
\affiliation{Institut f\"ur Kernphysik, TU Darmstadt, Schlossgartenstr. 2, 64289 Darmstadt, Germany} 

\author{G.~Hagen}
\affiliation{Physics Division, Oak Ridge National Laboratory,
Oak Ridge, TN 37831, USA} 
\affiliation{Department of Physics and Astronomy, University of Tennessee,
Knoxville, TN 37996, USA} 

\author{J.~D.~Holt}
\affiliation{TRIUMF 4004 Wesbrook Mall, Vancouver BC V6T 2A3, Canada} 

\author{G.~R.~Jansen}
\affiliation{National Center for Computational Sciences, Oak Ridge National
Laboratory, Oak Ridge, TN 37831, USA}
\affiliation{Physics Division, Oak Ridge National
Laboratory, Oak Ridge, TN 37831, USA}

\author{T.~Papenbrock}
\affiliation{Department of Physics and Astronomy, University of Tennessee,
Knoxville, TN 37996, USA} 
\affiliation{Physics Division, Oak Ridge National Laboratory,
Oak Ridge, TN 37831, USA} 

\author{R.~Roth}
\affiliation{Institut f\"ur Kernphysik, TU Darmstadt, Schlossgartenstr. 2, 64289 Darmstadt, Germany} 

\author{A.~Schwenk}
\affiliation{Institut f\"ur Kernphysik, TU Darmstadt, Schlossgartenstr. 2, 64289 Darmstadt, Germany} 
\affiliation{ExtreMe Matter Institute EMMI, GSI Helmholtzzentrum f\"ur Schwerionenforschung GmbH, 64291 Darmstadt, Germany}
\affiliation{Max-Planck-Institut f\"ur Kernphysik, Saupfercheckweg 1, 69117 Heidelberg, Germany}

\begin{abstract}

We link the structure of nuclei around $^{100}$Sn, the heaviest doubly
magic nucleus with equal neutron and proton numbers ($N=Z=50$), to
nucleon-nucleon ($NN$) and three-nucleon ($NNN$) forces constrained by
data of few-nucleon systems. Our results indicate that $^{100}$Sn is
doubly magic, and we predict its quadrupole collectivity. We present
precise computations of $^{101}$Sn based on three-particle--two-hole
excitations of $^{100}$Sn, and reproduce the small splitting between
the lowest $J^\pi=7/2^+$ and $5/2^+$ states.
Our results are consistent with the sparse available data.
\end{abstract}

\maketitle 

{\it Introduction} -- $^{100}$Sn is a nucleus of superlatives: It is
the heaviest self-conjugate ($N=Z=50$) nucleus~\cite{lewitowicz1994},
exhibits the largest strength in allowed $\beta$
decay~\cite{hinke2012}, is close to the proton
dripline~\cite{erler2012}, and is the endpoint of a region of nuclei
with enhanced $\alpha$ decays~\cite{liddick2006,seweryniak2006}. While
these properties make $^{100}$Sn the cornerstone of a most interesting
region of the nuclear chart, our understanding of this nucleus and its
neighbors is still rather limited, see \cite{faestermann2013} for a
review.  No data exist regarding the spectrum of $^{100}$Sn, and the
spin assignments for low-lying states in $^{101}$Sn are
controversial~\cite{seweryniak2007,darby2010}.  Likewise, the
evolution of collective observables towards neutron number $N\!=\!50$
is experimentally unclear at
present~\cite{banu2005,vaman2007,ekstrom2008,guastalla2013,bader2013,doornenbal2014}.
On the other hand, with naively expected shell closures for both
protons and neutrons, and the stabilizing effects of the Coulomb and
centrifugal barriers, $^{100}$Sn should be particularly suitable for a
reliable theoretical treatment.  

In this Letter, we calculate properties of $^{100}$Sn and neighboring
nuclei using realistic interactions between protons and neutrons.
This is in contrast to large-scale shell-model (LSSM) calculations
\cite{lipoglavsek2002,blazhev2004,boutachkov2010,brock2010,caurier2010}
in this region of the nuclear chart that employ $^{80}$Zr or $^{88}$Sr
cores and phenomenologically adjusted interactions based on the
$G$-matrix approach~\cite{hjorthjensen1995}. The strong nuclear force
is rooted in the fundamental theory of strong interactions, quantum
chromodynamics (QCD), and is manifested in dominant two-nucleon ($NN$)
forces and weaker but pivotal three-nucleon ($NNN$) forces between
protons and neutrons. Effective field theories (EFTs) of QCD provide
us with a systematically improvable low-momentum expansion of these
interactions~\cite{vankolck1999,epelbaum2009,machleidt2011}.  So far,
interactions derived from the EFT framework have been applied to light
and medium-mass nuclei
(see~\cite{navratil2009,barrett2013,hagen2014,hergert2016} for recent
reviews).

The extension of {\it ab initio} computations from
light~\cite{pieper2001,navratil2009,barrett2013} to heavier nuclei is
based on the development and application of quantum many-body methods
that exhibit a polynomial scaling in mass
number~\cite{mihaila2000b,dean2004,dickhoff2004,hagen2008,hagen2010b,tsukiyama2011,roth2012,hergert2013b,soma2013b,binder2013,lahde2014}.
Medium-mass and heavy nuclei can technically be computed with these
methods, but most interactions developed thus far considerably
overbind heavier nuclei~\cite{binder2013b}. In the quest for nuclear
interactions with more acceptable saturation
properties~\cite{hebeler2011,ekstrom2013,ekstrom2015,ekstrom2017}, one
interaction labeled 1.8/2.0(EM) has emerged that describes binding
energies, two-neutron separation energies, and the first $2^+$ excited
state in nuclei up to neutron-rich nickel isotopes remarkably well,
while charge radii are too
small~\cite{hagen2015,hagen2016,hagen2016b,simonis2016,garciaruiz2016,simonis2017}. It
is primarily this interaction from Ref.~\cite{hebeler2011} that we
will employ in the computation of $^{100}$Sn and its neighbors.

This Letter is organized as follows. First, we briefly describe the
Hamiltonian, the employed model spaces, and computational
methods. Then, we validate the interactions in the tin region by
computing known binding energies and level splittings, and address
method uncertainties by employing the coupled-cluster
method~\cite{bartlett2007,hagen2014} and the 
valence-space in-medium similarity-renormalization-group method
(VS-IMSRG)~\cite{stroberg2017} coupled with the importance-truncated
large-scale shell model~\cite{stumpf2016}. Finally, we present results
for the structure of the lightest isotopes of tin.

{\it Hamiltonian and model space} -- We employ the intrinsic Hamiltonian
\begin{equation}
  \label{intham}
  H = \sum_{i<j}\left({({\bf p}_i-{\bf p}_j)^2\over 2mA} + V
    _{NN}^{(i,j)}\right) + \sum_{ i<j<k}V_{NNN}^{(i,j,k)} ,
\end{equation}
where the $NN$ and $NNN$ potentials ($V_{NN}$ and $V_{NNN}$,
respectively) are the interactions 1.8/2.0(EM), 2.0/2.0(EM),
2.2/2.0(EM), and 2.0/2.0(PWA) from Ref.~\cite{hebeler2011}. These
interactions result from a similarity-renormalization-group
(SRG)~\cite{bogner2007} evolution of the chiral $NN$ interaction of
\cite{entem2003} to cutoffs $\lambda=1.8,2.0,$ and
$2.2$~fm$^{-1}$, respectively. The $NNN$ potential is not evolved but
rather taken as the leading $NNN$ forces from chiral
EFT~\cite{vankolck1994,epelbaum2002,hebeler2015b} and has a cutoff
$\Lambda_{NNN}=2.0$~fm$^{-1}$. The two low-energy constants of the
short-range part of the $NNN$ forces are adjusted to binding energy
of the triton and the radius of the $\alpha$ particle, following Ref.~\cite{nogga2004}. 
These interactions are quite
soft (due to the relatively small cutoffs), which allows us to achieve
reasonably well converged binding energies and spectra in nuclei up to
neutron-rich $^{78}$Ni~\cite{hagen2016b,simonis2017}, and in the
neutron-deficient tin isotopes considered in this work.

\begin{figure}[t]
  \includegraphics[width=1.0\columnwidth]{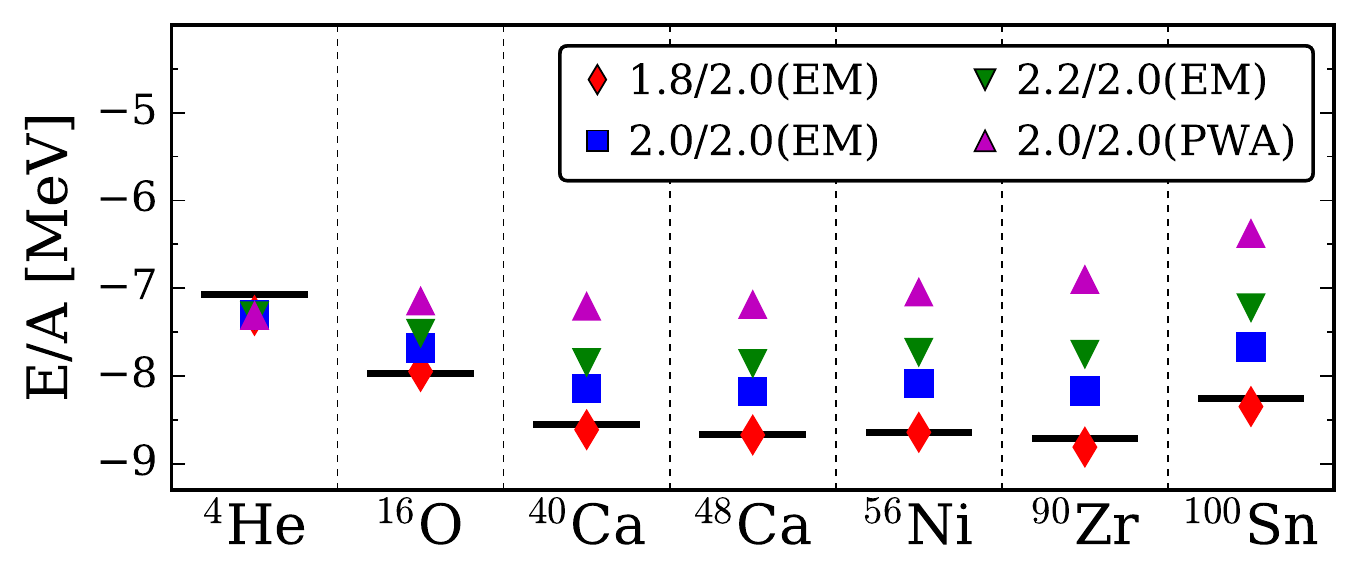}
  \caption{(Color online) Ground-state energies per nucleon $E/A$ for
    selected closed-shell nuclei computed with the closed-shell
    IMSRG~\cite{tsukiyama2011} using the interactions of
    Ref.~\cite{hebeler2011} in comparison with experiment (black
    horizontal lines).}
  \label{BE_A}
\end{figure}

Figure \ref{BE_A} shows the computed ground-state energies per nucleon
for $^{4}$He, $^{16}$O, $^{40,48}$Ca, $^{56}$Ni, $^{90}$Zr, and
$^{100}$Sn with the single-reference
IMSRG~\cite{tsukiyama2011,hergert2013b}.  The 1.8/2.0(EM) interaction
consistently yields the best agreement with data.  Presently, it is
unclear what distinguishes this interaction from the other similarly
obtained interactions; however this soft interaction puts us in a
fortuitous situation to make theoretical predictions (albeit without
rigorous uncertainty quantification) for binding energies and spectra
in nuclei as heavy as $^{100}$Sn.

Coupled-cluster calculations use a Hartree-Fock basis constructed from
a harmonic-oscillator basis of up to 15 major oscillator shells. For
VS-IMSRG we use a similar basis, except that the Hartree-Fock reference
is constructed with respect to an ensemble state above the $^{80}$Zr
core following Ref.~\cite{stroberg2017}.  All calculations are performed at
oscillator frequencies in the range $\hbar\omega = 12-16$~MeV, which
include the minimum in energy for the largest model space we
consider. We use the normal-ordered two-body approximation
\cite{hagen2007a,roth2012,binder2013b} for the $NNN$ interaction with an
additional energy cut on three-body matrix elements $e_1+e_2+e_3\leq E_{\rm 3 max}$.
When $E_{\rm 3 max}$ is increased from 16 to 18, the binding energy of $^{100}$Sn
 changes by 2\% for the hardest interaction 2.0/2.0(PWA), while
for the softest interaction, 1.8/2.0(EM), the change is less than 1\%.

{\it Method } -- The coupled-cluster method is an ideal tool to
compute doubly magic nuclei and their
neighbors~\cite{kuemmel1978,mihaila2000b,dean2004,wloch2005,hagen2008,hagen2010b,roth2011a,binder2013,hagen2014}.
This method computes the similarity transform
$\overline{H}\equiv \exp{(-T)}H_N\exp{(T)}$ of the Hamiltonian $H_N$,
obtained by normal ordering the free-space Hamiltonian~(\ref{intham})
with respect to the closed-shell Hartree-Fock reference of
$^{100}$Sn. The cluster operator $T$ includes particle-hole
excitations and is truncated at the coupled-cluster singles-doubles
(CCSD) level. Usually CCSD accounts for about 90\% of the correlation
energy (i.e., the energy beyond Hartree Fock)~\cite{bartlett2007}. For
a higher precision of the ground-state energy, we include triples
excitations of the cluster operator $T$ perturbatively within the
$\Lambda$-CCSD(T) method \cite{taube2008}. Excited states in
$^{100}$Sn are computed with an equation-of-motion (EOM) method
including 3$p$-3$h$ corrections via a generalization of the ground
state $\Lambda$-CCSD(T) approximations to excited states with
EOM-CCSD(T) ~\cite{watts1995}. The neighboring nuclei $^{101,102}$Sn
are computed as one- and two-particle attached
states~\cite{gour2005,jansen2011,jansen2012} of the $^{100}$Sn
similarity transformed Hamiltonian $\overline{H}$.  The two-particle
attached states of $^{102}$Sn are truncated at the $3p$-$1h$ level,
while the particle-attached states of $^{101}$Sn are computed at the
$2p$-$1h$ level with perturbative $3p$-$2h$ corrections included
(described below). Further details of the coupled-cluster approach to
nuclei are presented in a recent review~\cite{hagen2014}.

We briefly describe our new approach to include perturbative $3p$-$2h$
corrections to the particle-attached states of $^{101}$Sn.
Generalizing the completely renormalized (CR) EOM-CCSD(T)
approximation from quantum chemistry~\cite{kowalski2000,piecuch2002}
and nuclear physics~\cite{kowalski2004,wloch2005,binder2013} to
particle-attached excited states yields the correction
\begin{eqnarray}
\delta \omega_{\nu}^{3p\text{-}2h} = \sum_{i<j}\sum_{a<b<c} \mathcal{L}_{\nu ,ij}^{abc}\mathcal{R}_{\nu, ij}^{abc}\mathcal{M}_{\nu, ij}^{abc} .\label{eq:ecorrection}
\end{eqnarray}
Here $\nu$ denotes the state of interest, $i,j$ (and $a,b,c$) are
occupied (and unoccupied) orbitals in the $^{100}$Sn reference
$\vert \Phi\rangle$, $\mathcal{L}_{\nu}$ and $\mathcal{M}_{\nu}$
represent the left and right $3p$-$2h$ moments 
\begin{eqnarray}
\mathcal{L}_{\nu ,ij}^{abc} = \langle \Phi | L_{\nu}^{2p\text{-}1h} \overline{H} | \Phi_{ij}^{abc} \rangle \ , \ 
\mathcal{M}_{\nu ,ij}^{abc} = \langle \Phi_{ij}^{abc} | \overline{H} R_{\nu}^{2p\text{-}1h} | \Phi \rangle \ ,\notag 
\end{eqnarray}
$| \Phi_{ij}^{abc} \rangle$ are $3p$-$2h$ excited states, and
$\mathcal{R}_{\nu}$ is the resolvent
\begin{eqnarray}
\mathcal{R}_{\nu ,ij}^{abc} = \langle \Phi_{ij}^{abc} | (\omega_{\nu}^{2p\text{-}1h} - \overline{H})^{-1} | \Phi_{ij}^{abc} \rangle . \label{eq:resolvent}
\end{eqnarray}
Here $\omega_{\nu}^{2p\text{-}1h}$ is the $2p$-$1h$ energy
corresponding to the states $L_{\nu}^{2p\text{-}1h}$ and
$R_{\nu}^{2p\text{-}1h}$ of $^{101}$Sn.  We draw the reader's
attention to the similar structure between the bi-variational
expression~(\ref{eq:ecorrection}) and second-order perturbation
theory. This method is the completely renormalized particle-attached
equation-of-motion (CR-PA-EOM). In our results for $^{101}$Sn, we used
three different approximations (labeled A,B,C) for the energy
denominator in Eq.~(\ref{eq:resolvent}).  Approximation A uses in
place of $\overline{H}$ the Hartree-Fock single-particle energies,
approximation B uses the one-body part of $\overline{H}$, and
approximation C uses both the one- and two-body parts of
$\overline{H}$. Thus, approximation C is the most complete choice for
the resolvent and most accurately approximates the full
calculation~\cite{wloch2005}.
 
The IMSRG and its VS-IMSRG variant are effective tools for computing
doubly magic nuclei and for constructing valence-space interactions
from $NN$ and $NNN$ interactions that can be subsequently diagonalized
using shell-model
techniques~\cite{tsukiyama2011,tsukiyama2012,bogner2014,hergert2016,
  stroberg2016,stroberg2017}. These methods also rely on similarity
transformations $\overline{H}\equiv \exp{(\Omega)}H_N\exp{(-\Omega)}$
where $H_N$ is the normal-ordered Hamiltonian with respect to the
ensemble reference of each target nucleus. For nuclei in the
$^{100}$Sn region, the VS-IMSRG yields an anti-Hermitian $\Omega$,
truncated at the one- and two-body level, which decouples the major
oscillator shell above $^{80}$Zr. The ensuing large-scale eigenvalue
problem is solved via the importance-truncated shell
model~\cite{stumpf2016}.

\begin{figure*}[t]
  \input{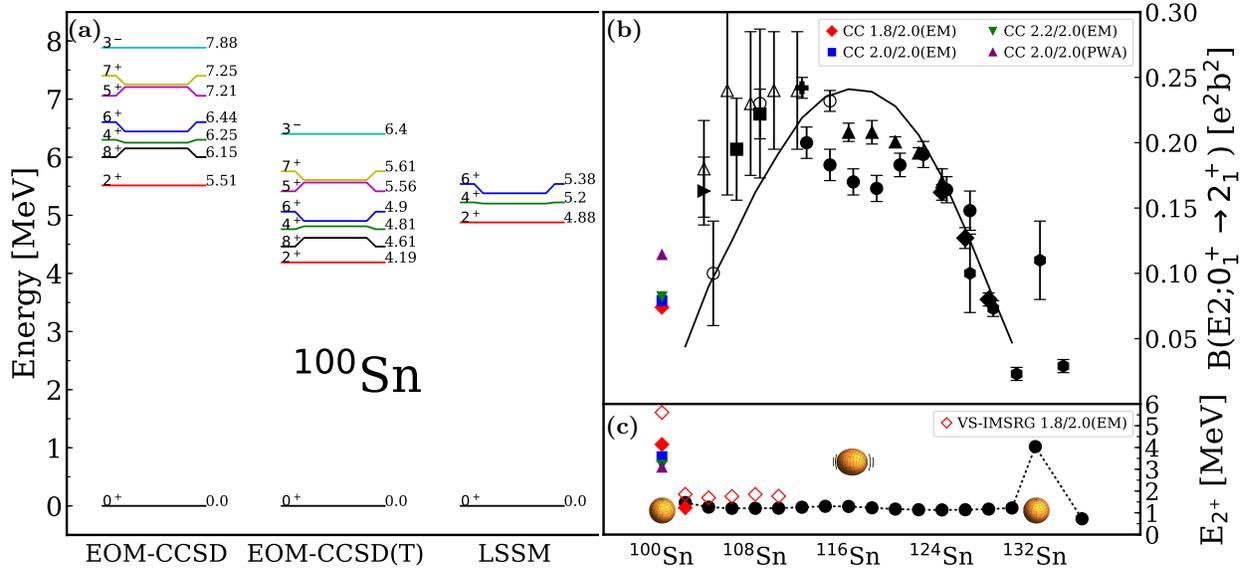}
  \caption{(Color online) Panel (a) shows low-lying states in
    $^{100}$Sn computed with the chiral interaction 1.8/2.0(EM) 
    in the EOM-CCSD and EOM-CCSD(T) approximations
    and compared to LSSM calculations based on phenomenological interactions
    \cite{faestermann2013}. Panel (b) shows the EOM-CCSD
    results for the $B(E2)$ transition strength in $^{100}$Sn with the 
    interactions as indicated, and the experimental data for all other even tin
    isotopes. Panel (c) shows the energy of the $J^\pi=2_1^+$ states
    in even tin isotopes, with coupled-cluster results for
    $^{100,102}$Sn [labelled as in panel (b)] and VS-IMSRG results for $^{100-110}$Sn
    with interactions as indicated, and data for $^{102-132}$Sn (black
    circles). }
  \label{Sn100combo}
\end{figure*}

{\it Results} -- Results for $^{100}$Sn are shown in
Fig.~\ref{Sn100combo}.  Panel (a) shows the low-lying states in
$^{100}$Sn computed in the EOM-CCSD and EOM-CCSD(T) approximations
with the 1.8/2.0(EM) interaction.  We also show the phenomenological
LSSM results of Ref.~\cite{hinke2012}. The relatively large excitation
gap of about 4~MeV is consistent with $^{100}$Sn being doubly magic, a
finding which is---to our knowledge---qualitatively ubiquitous in all
previous theoretical investigations.  Panel (b) shows our EOM-CCSD
predictions for the $B(E2)$ in $^{100}$Sn for the 1.8/2.0(EM),
2.0/2.0(EM), 2.2/2.0(EM), and 2.0/2.0(PWA) interactions together with
the experimental $B(E2)$ values for the isotopes
$^{104-132}$Sn~\cite{banu2005,vaman2007,ekstrom2008,guastalla2013,bader2013,doornenbal2014}.
Our computed $B(E2)$ values are similar in size to that of $^{132}$Sn
and consistent with $^{100}$Sn being doubly magic. They also fall
within expectations from phenomenological shell-model
calculations~\cite{faestermann2013} and from extrapolations of data in
light tin isotopes. Panel (c) shows CC results for 1.8/2.0(EM),
2.0/2.0(EM), 2.2/2.0(EM), and 2.0/2.0(PWA) interactions and the
VS-IMSRG result for the 1.8/2.0(EM) interaction for the energy of the
first $J^\pi=2_1^+$ state in light even isotopes $^{100-110}$Sn, and
data for $^{102-132}$Sn. The $B(E2)$ values computed are consistent
with the computed excitation energies of the first $J^\pi=2_1^+$, in
the sense that, for a given interaction, the larger the energy, the
smaller the $B(E2)$ value.  We note that despite the consistency with
experiment, the 1.8/2.0(EM) interaction produces radii that are too
small, and this would certainly affect the $B(E2)$. The systematic
trend of known and computed $2^+_1$ energies in the tin isotopes again
suggests that $^{100}$Sn is doubly magic. In $^{100}$Sn, this energy
is similar to that of the doubly magic nucleus
$^{132}$Sn~\cite{bjornstad1980,jones2010}. The VS-IMSRG result for the
$2^+_1$ state in $^{100}$Sn is about $1.5$~MeV higher than the
EOM-CCSD(T) result, but close to the similarly approximated EOM-CCSD
result shown in panel (a).  Using the discrepancy between methods as
an estimate of the uncertainty in the many-body method, our results
for the energy of $2^+_1$ state in $^{102\text{-}110}$Sn are
consistent with the data.

\begin{figure*}[t!]
  \input{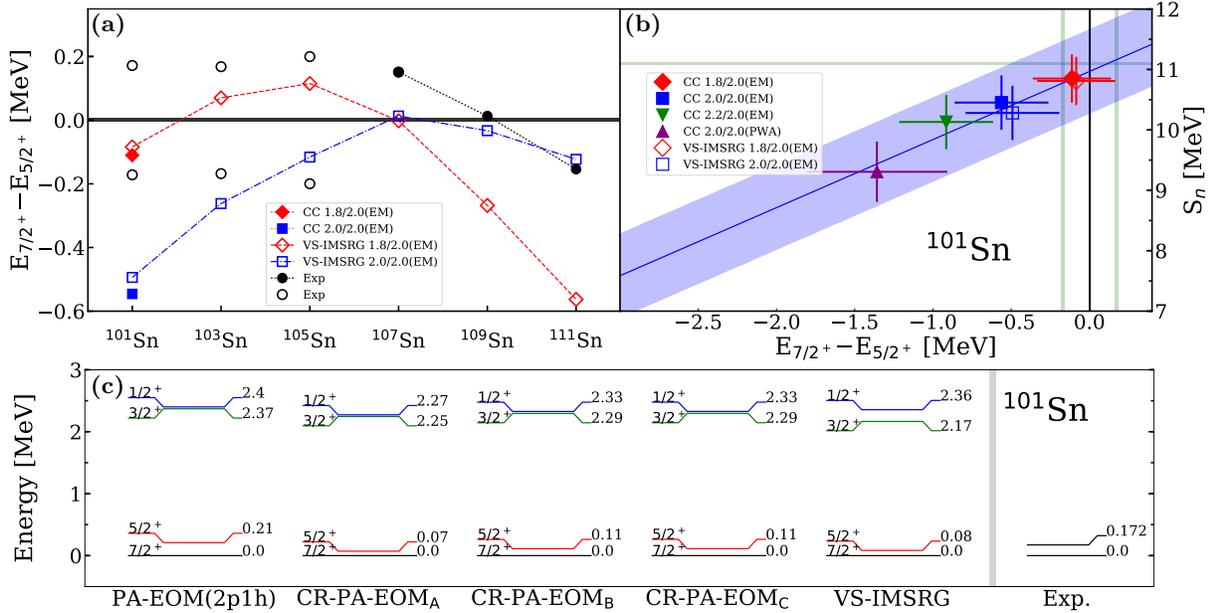}
  \caption{(Color online) Panel (a) shows the energy splitting between
    the lowest $J^\pi={7/2}^+$ and ${5/2}^+$ states in light odd-mass
    tin isotopes.  Data are black circles for isotopes with definite
    (full) and tentative spin assignments (open symbols). Theoretical
    results are from coupled cluster (full) and VS-IMSRG
    (open symbols) for the interactions as indicated. Panel (b)
    shows the correlation between the neutron separation energy $S_n$
    and the energy splitting between the lowest $J^\pi={7/2}^+$ and
    ${5/2}^+$ states for the interactions and computational method as
    labeled (symbols with error bars and an encompassing blue
    uncertainty band) compared to data (vertical and horizontal green
    lines).  Panel (c) shows the low-lying levels in $^{101}$Sn based
    on the chiral interaction 1.8/2.0(EM) computed with various
    coupled-cluster equation-of-motion (EOM) methods, the VS-IMSRG,
    and compared to data.}
    \label{Sn101combo}
\end{figure*}

In $^{101}$Sn, the two lowest states are separated by only
172~keV~\cite{seweryniak2007,darby2010}. Observation of $^{105}$Te
$\alpha$ decay in coincidence with the 172~keV $\gamma$ line indicates
that the dominant $\alpha$ decay of $^{105}$Te is to the first excited
state in $^{101}$Sn, implying that these states have identical
spins~\cite{darby2010}. We recall that the lowest two states in the
odd isotopes $^{105-113}$Te and $^{101-105}$Sn are only about 0.2~MeV
apart and lack definite spin assignments. In tin, this near degeneracy
between the $J^\pi=5/2^+$ and $7/2^+$ states persists up to
$^{111}$Sn, and the ground-state spin changes from $J^\pi=5/2^+$ in
$^{107}$Sn~\cite{cerizza2016} and $^{109}$Sn to $J^\pi=7/2^+$ in
$^{111}$Sn.  The level ordering in $^{101}$Sn to $^{105}$Sn between
the $J^\pi=5/2^+$ and $7/2^+$ states is not known.  This is reflected
in panel (a) of Fig.~\ref{Sn101combo} which compares available data
(full and open black points for definite and tentative spins
assignments, respectively) with CC and VS-IMSRG predictions for the
energy splitting in odd tin isotopes using the interactions
1.8/2.0(EM) and 2.0/2.0(EM). Both interactions yield a small splitting
between the $J^\pi=5/2^+$ and $7/2^+$ states, but they differ on its
precise size and sign. Panel (b) of Fig.~\ref{Sn101combo} plots the
calculated energy splitting between the $J^\pi=5/2^+$ and $7/2^+$
states versus the neutron separation energy of $^{101}$Sn computed
with the CR-PA-EOM using the 1.8/2.0(EM), 2.0/2.0(EM), 2.2/2.0(EM),
and 2.0/2.0(PWA) interactions. Also shown are estimated uncertainties
due to finite model-space sizes and the employed methods, and a blue
(diagonal) band encompassing these uncertainties (see
Ref.~\cite{hagen2015} for details). The horizontal and vertical green
lines indicate experimental data. The intersection of the blue
diagonal band with the precisely known neutron separation energy $S_n$
yields one estimate of the systematic uncertainty for the energy
splitting between the $J^\pi=5/2^+$ and $7/2^+$ states in
$^{101}$Sn. Clearly, theory is not sufficiently precise to make a
definite prediction for the ground-state spin of $^{101}$Sn as the
predicted range for the energy splitting can support either
$J^\pi=5/2^+$ or $7/2^+$ as the ground state. Again, the 1.8/2.0(EM)
interaction is closest to data. Panel (c) of Fig.~\ref{Sn101combo}
shows the lowest states in $^{101}$Sn, computed with the $2p$-$1h$
particle-attached EOM-CC method, the CR-PA-EOM developed in this work,
and the VS-IMSRG for the 1.8/2.0(EM) interaction. We find that for
this interaction, the different methods agree on the level ordering,
and the energy splitting varies by at most $140$~keV.  While the
upcoming measurements will yield definite spin
assignments~\cite{garciaruiz2016b}, getting theory to a level where
such fine details can be unambiguously resolved will require more
work.

\begin{figure}[t]
  \includegraphics[width=0.5\textwidth]{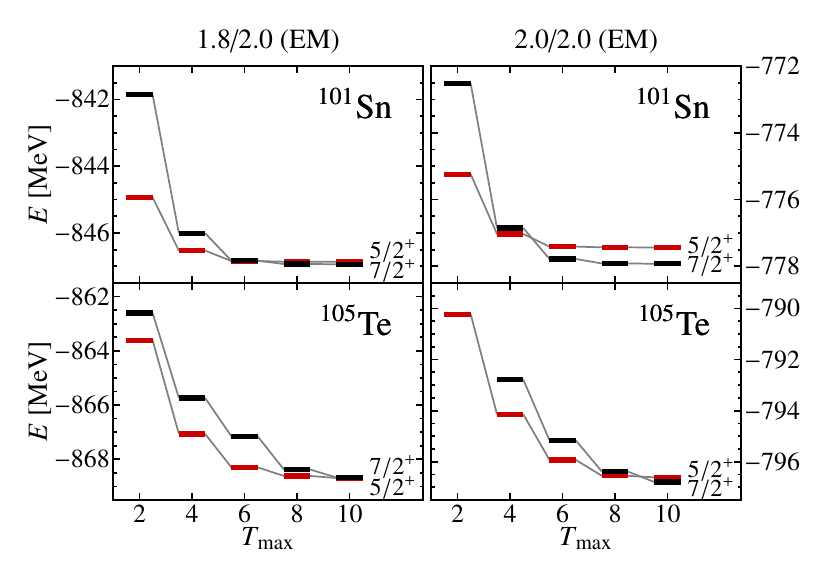}
  \caption{(Color online) Ground- and first excited states in
    $^{101}$Sn and $^{105}$Te obtained in VS-IMSRG for the 1.8/2.0(EM) and 2.0/2.0(EM)
    interactions, with spins $J^\pi=5/2^+$ (red) and $J^\pi=7/2^+$
    (black).}
  \label{te_sn_combi}
\end{figure}

Figure~\ref{te_sn_combi} shows the convergence of the ${5/2}^+$ and
${7/2}^+$ states in $^{101}$Sn and $^{105}$Te with the number of
particle-hole excitations ($T_{\rm max}$) in the importance-truncated
large-scale shell-model calculations using the 1.8/2.0(EM) and
2.0/2.0(EM) interactions. In both, $^{101}$Sn and $^{105}$Te, we
obtain nearly degenerate $J^\pi={5/2}^+$ and ${7/2}^+$ states
consistent with data. 

{\it Conclusions and Outlook} -- Our computations demonstrated that
tin nuclei can be described by $NN$ and $NNN$ interactions constrained
by few-body data. We found that $^{100}$Sn is doubly magic and
presented results and predictions for its structure and low-lying
collectivity. For an increased precision of excited states in
$^{101}$Sn, we developed a method that includes
three-particle-two-hole excitations in our coupled-cluster
calculations.  One interaction reproduced both binding energies and
the near degeneracy between the lowest $J^\pi={5/2}^+$ and ${7/2}^+$
states in the odd-mass isotopes $^{101-111}$Sn and $^{105}$Te. This
work opens the avenue for reliable calculations of even heavier nuclei
based on $NN$ and $NNN$ interactions.

\begin{acknowledgments}
  We thank K. Hebeler for providing us with matrix elements in Jacobi
  coordinates for the $NNN$ interaction at next-to-next-to-leading
  order. This work was supported by the Office of Nuclear Physics,
  U.S. Department of Energy, under Grants DE-FG02-96ER40963,
  DE-SC0008499 (NUCLEI SciDAC collaboration), the Field Work Proposal
  ERKBP57 at Oak Ridge National Laboratory (ORNL), by the ERC Grant
  No.~307986 STRONGINT, the BMBF under Contract No.~05P15RDFN1, the
  DFG under Grant SFB~1245 DFG, the National Research Council Canada,
  and NSERC. Computer time was provided by the Innovative and Novel
  Computational Impact on Theory and Experiment (INCITE) program. This
  research used resources of the Oak Ridge Leadership Computing
  Facility located at ORNL, which is supported by the Office of
  Science of the Department of Energy under Contract No.
  DE-AC05-00OR22725. Computations were also performed at the
  LICHTENBERG high performance computer of the TU Darmstadt, the
  J\"ulich Supercomputing Center (JURECA), the LOEWE-CSC Frankfurt,
  and at the Max-Planck-Institute for Nuclear Physics.
\end{acknowledgments}

\bibliography{refs}

\end{document}